\documentclass[
    ,final            % use final for the camera ready runs
  ]
  {aipproc}

\layoutstyle{8x11double}

\newcommand{\approxlt}{\mbox{$\;^{<}\hspace{-0.24cm}_{\sim}\;$}}
\newcommand{\approxgt}{\mbox{$\;^{>}\hspace{-0.24cm}_{\sim}\;$}}

\begin{document}

\title{A Search for Single Radio Pulses and Bursts from Southern AXPs}

\classification{97.60.Jd, 98.70.Qy}
\keywords      {neutron stars, AXPs, magnetars}

\author{F. Crawford}{
  address={Department of Physics and Astronomy, Franklin \& Marshall 
College, Lancaster, PA 17604, USA}
}

\author{J. W. T. Hessels}{
  address={Astronomical Institute ``Anton Pannekoek,'' University of 
Amsterdam, Kruislaan 403, 1098 SJ Amsterdam, The Netherlands}
}

\author{V. M. Kaspi}{
  address={Department of Physics, McGill University, Montreal, 
QC H3A 2T8, Canada}
%  ,altaddress={} % additional visiting address
}

\begin{abstract}
We observed four southern AXPs in 1999 near 1400 MHz with the Parkes
64-m radio telescope to search for periodic radio emission. No Fourier
candidates were discovered in the initial analysis, but the recent
radio activity observed for the AXP XTE J1810$-$197 has prompted us to
revisit these data to search for single radio pulses and bursts. The
data were searched for both persistent and bursting radio emission at
a wide range of dispersion measures, but no detections of either kind
were made. These results further weaken the proposed link between
rotating radio transient sources and magnetars. However, continued
radio searches of these and other AXPs at different epochs are
warranted given the transient nature of the radio emission seen from
XTE J1810$-$197, which until very recently was the only known
radio-emitting AXP.
\end{abstract}

\maketitle

\section{Introduction and Motivation}

The detection of pulsed radio emission from the anomalous X-ray pulsar
(AXP) XTE J1810$-$197 in 2006 \cite{crh+06}, and more recently from a
second AXP, 1E 1547.0$-$5408 \cite{crh+07}, has renewed interest in
searching for radio emission from these objects. In both of these
cases, the radio activity is believed to be connected to the X-ray
variability of the sources and is transient in nature (or at least
highly variable). Given this transient behavior and that both
persistent periodic emission and single pulses were detected from both
AXPs, renewed searches of archival radio search data of AXPs at
different epochs may reveal previously undetected radio signals from
these sources.

\section{Observations and Analysis}

Three southern AXPs and one AXP candidate were observed in July and
August 1999 with the Parkes 64-m radio telescope. The four targets
observed were:

\begin{itemize}

\item 1E 1048.1$-$5937
\item AX J1845$-$0258 (AXP candidate only)
\item 1E 1841$-$045
\item 1RXS J170849.0$-$400910

\end{itemize}

All observations were conducted with the center beam of the multibeam
receiver \cite{swb+96} at a center frequency of 1374 MHz. 288 MHz of
bandwidth was split into 96 frequency channels. This is the same
observing setup as was used for the Parkes Multibeam Survey
\cite{mlc+01} and the AXP search observations reported by
\cite{bri+06}. None of the four targets was in a state of X-ray
outburst at the time of observation.

The data were processed using the
PRESTO\footnote{http://www.cv.nrao.edu/$\sim$sransom/presto} suite of
pulsar analysis tools \citep{r01,rem02}.  First, the raw data were
excised of radio frequency interference (RFI). This is particularly
important given the slow rotation rates of the AXPs and the pernicious
effect of RFI at low modulation frequencies. Typically, 15-25\% of the
frequency channels and $\sim 5$\% of the integration time were
discarded in this process. Note that this data excision increases our
sensitivity limits by $\sim 10$\% and is not included in the estimates
presented in Table \ref{tbl-1}. Standard Fourier searches of the data
were previously reported \citep{cpk+02}, but no signals were confirmed
in that analysis.

In both the folding search and single pulse analysis reported here,
the data were dedispersed at a wide range of dispersion measures
(DMs). The DMs ranged from 0 to 4000-8000 pc cm$^{-3}$, depending on
the spin period of the AXP. The dedispersed time series were searched
for both persistent periodic emission and single pulses and bursts.

To detect periodic emission from these sources, the data were divided
into 9 MHz subbands, and each subband was dedispersed at a trial DM of
500 pc cm$^{-3}$. Each subband was then folded at the known neutron
star spin period using the X-ray timing ephemeris (where
available). In the case of AX J1845$-$0258, where no ephemeris was
available, the discovery period was used. The full range of trial DMs
was created by shifting the subbands with respect to each other in
pulse phase. We also folded the data allowing for a search in period
$\pm 5$-10 ms from the predicted period (depending on the spin
period).  Additionally, separate folds were made for each target using
overlapping separate shorter segments of the full observation; each of
these segments was 20\% of the full observation length and started at
intervals of 10\% of the data length. This accounted for possible
scintillation, strong transient RFI near the AXP spin period, and
pulse strength variability on time-scales shorter than the observation
length.

For the single pulse search, the raw data were again dedispersed at a
set of trial DMs ranging from 0 to 4000-8000 pc cm$^{-3}$. Using the
0.25 ms native sampling of the data, each dedispersed time series was
searched for candidate signals having a signal-to-noise ratio greater
than 6.5. This threshold was chosen to avoid confusion with the RFI
background. To maintain sensitivity to pulses of width greater than
0.25 ms, a matched filter was employed using a boxcar function of
varying width (ranging from 1 to 30 samples). To enhance sensitivity
to even longer pulses and bursts, the dedispersed time series was
downsampled by combining the original samples into contiguous blocks
of 2, 4, 8, 16, and 32 samples, and the same boxcar filtering was then
used. This provided sensitivity to bursts with durations up to 240 ms.

\section{Results and Conclusions}

We found no convincing radio signals in either the folding or single
pulse searches. The derived upper limits on the radio emission from
our AXP targets are presented in Table \ref{tbl-1}. These are the most
stringent radio upper limits to date for these sources. The estimated
1400 MHz luminosity limits on the periodic radio emission ($\approxlt
1$ mJy kpc$^{2}$) are 2-3 times lower than those established for XTE
J1810$-$197 prior to outburst. However, it is still conceivable that
weak radio pulses are being emitted, but that they are below our
detection threshold.

The luminosity limits presented here for the periodic emission are
lower by about two orders of magnitude than the 1400 MHz luminosity of
the periodic pulsed radio emission from XTE J1810$-$197 soon after the
radio emission was first detected ($\sim 80$ mJy kpc$^{2}$)
\citep{crh+06}. Thus we would expect to be able to easily detect
comparably strong radio emission if it were beamed toward us.

Our luminosity limits on single radio pulses from our sources range
from 22 to 69 Jy kpc$^{2}$ in the most conservative case, which is
below the 1400 MHz luminosity of $\approxgt 100$ Jy kpc$^{2}$ derived
from the pulse strengths reported for XTE J1810$-$197 in its radio
discovery paper \citep{crh+06}. Since single pulses were detected from
almost every rotation of XTE J1810$-$197, it is likely that we would
have detected a large number of comparable pulses during our
observations if such pulses were beamed toward us.

Our non-detection of single pulses further weakens the hypothesis that
rotating radio transients (RRATs) \citep{mlm+06} and magnetars are
linked. This has been weakened by two other recent results. First, the
X-ray detection of the RRAT J1819$-$1458 shows that its emission is
more typical of middle-aged pulsars than it is of magnetars
\citep{rbg+06}. Second, the nearby, rotation-powered pulsar PSR
B0656+54 would probably have been identified as an RRAT if it were
farther away \citep{wsr+06}.

We conclude from our results that any periodic or bursting radio
emission from the four target AXPs is either very weak (below our
detection thresholds), not beamed toward us, or non-existent or
sporadic at the epoch of observation. This last possibility is
suggested by the connection between the X-ray and radio activity
observed for the two known radio-emitting magnetars to date.
Continued radio searches of AXPs are therefore warranted given the
apparent transient nature of the radio emission. Further details of
this work and a more complete discussion of the results are presented
in a recent journal article \cite{chk07}.

%%%%%%%%%%%%%%%%%%%%%%%%%%%%%%%%%%%%%%%%%%%%
%% TABLES:
%%
%% 1. \begin{table}[pos], the optional pos argument can be used to specify
%%	which float areas this float is allowed to migrate (default is tbp).
%%
%% 2. \tablehead command is provided, used as
%% 	\tablehead{cols}{h-pos}{v-pos}{heading text}, with 
%%	cols  specifying the numbers of columns the heading text should span,
%%	h-pos defining the horizontal positioning of the text of 
%%	      the column(s), e.g., l, r, c, or p{...},
%%	v-pos containing either t, c, or b to denote the vertical placement
%%	      of the text in relation to other cells of that row (this is
%%	      only relevant if the heading text consists of more than one 
%%	      line.
%%
%% 3. Heading text can be split vertically by using "\\" to denote the
%%	the line breaks.
%% 
%% 4. \tablenote{text} command produce a note to a table with text  
%%	appearing below the table.
%%
%% 5. The environment table* is not supported. Tables that need to
%%	span both columns are **automatically** recognized in two column mode.
%%
%%%%%%%%%%%%%%%%%%%%%%%%%%%%%%%%%%%
%% SAMPLE TABLE
%%
%% Shows the use of \tablehead and \tablenote
%% macros
%%%%%%%%%%%%%%%%%%%%%%%%%%%%%%%%%%%%%%%%%%%%

\begin{table}
\begin{tabular}{lcccc}
\hline
  & \tablehead{1}{r}{b}{1E 1048.1$-$5937}
  & \tablehead{1}{r}{b}{AX J1845$-$0258\tablenote{AXP candidate only}}
  & \tablehead{1}{r}{b}{1E 1841$-$045}
  & \tablehead{1}{r}{b}{1RXS J170849.0$-$400910}   \\
\hline
Spin period (s)     & 6.45 & 6.97 & 11.77 & 11.00 \\
Ephemeris reference & \cite{kgc+01} & \cite{tkk+98}\tablenote{No period derivative available} & \cite{gvd99} & \cite{gk02} \\
Galactic longitude, latitude (deg) & 288.26, $-$0.52 & 29.52, 0.07 & 27.39, $-$0.01 & 346.47, 0.03 \\
$T_{\rm sky}$ (K)\tablenote{1374 MHz sky temperature estimated from \cite{hss+82} assuming a spectral index of $-2.6$} & 9.1 & 12.3 & 13.2 & 16.3 \\
Observation MJD & 51378 & 51391 & 51382 & 51379 \\
Observation date & 1999 Jul 19 & 1999 Aug 1 & 1999 Jul 23 & 1999 Jul 20 \\
$S_{1400}$ (mJy)\tablenote{1400 MHz flux density limit on pulsed emission estimated using the modified radiometer equation and an assumed duty cycle of 2.7\%} & $\approxlt 0.02$ & $\approxlt 0.02$ & $\approxlt 0.02$ & $\approxlt 0.02$ \\
$S_{1400}$ single (mJy)\tablenote{Range of single-pulse 1400 MHz flux limits  for pulse time-scales 0.25-240 ms} & $\approxlt 875$-50 & $\approxlt 975$-60 & $\approxlt 1000$-60 & $\approxlt 1085$-65 \\
Distance (kpc)\tablenote{Taken from \cite{bri+06}. Question marks indicate significant uncertainty in the value} & $\sim 5$? & $\sim 8$? & $\sim 7$ & $\sim 8$? \\
$L_{1400}$ (mJy kpc$^{2}$)\tablenote{1400 MHz luminosity limit on pulsed emission, assuming a 1 sr beaming fraction} & $\approxlt 0.5$ & $\approxlt 1.3$ & $\approxlt 1.0$ & $\approxlt 1.3$ \\
$L_{1400}$ single (Jy kpc$^{2}$)\tablenote{Range of 1400 MHz luminosity limits on single pulses for pulse time-scales 0.25-240 ms} & $\approxlt 22$-1.3 & $\approxlt 62$-3.7 & $\approxlt 49$-2.9 & $\approxlt 69$-4.1 \\
\hline
\end{tabular}
\caption{Radio Search Parameters and Results}
\label{tbl-1}
\end{table}

%%%%%%%%%%%%%%%%%%%%%%%%%%%%%%%%%%%%%%%%%%%%%%%%
%% BACKMATTER
%%%%%%%%%%%%%%%%%%%%%%%%%%%%%%%%%%%%%%%%%%%%%%%%

%\begin{theacknowledgments}
%Thank you.
%\end{theacknowledgments}

%% The code below assumes that BibTeX is used. Compliant BibTex styles
%% are aipproc (for use with natbib) and aipprocl (if natbib is missing
%% at the site).
%%
%% Please run "bibtex \jobname" to obtain the bibliography and 
%% then re-run LaTeX twice to fix the references!
%%
%% When referring to citations in the text, in quare brackets [] show
%% the number in order of appearance. References in the References
%% section are listed in the same numerical order.
%%%%%%%%%%%%%%%%%%%%%%%%%%%%%%%%%%%%%%%%%%%%%%%%%

\bibliographystyle{aipproc}   % if natbib is available


\bibliography{crawford_fronefield_1}

%%%%%%%%%%%%%%%%%%%%%%%%%%%%%%%%%%%%%%%%%%%%%%%%%
%% If the bibliography is
%% produced without BibTeX, comment out the above lines, use
%% \begin{thebibliography}{widest-label} environment to hold 
%% the list of references and 
%% \bibitem{label} command to start a bibliographical entry having
%% the "label" for use in \cite commands.

\end{document}